\documentstyle[aps,prb]{revtex}
\makeatletter
\@addtoreset{equation}{section}
\makeatother
\newcommand{\bea}{\begin{equation}}
\newcommand{\eea}{\end{equation}}
\newtheorem{theorem}{Theorem}[section]
\newtheorem{lemma}{Lemma}[section]
\newtheorem{corrolary}{Corrolary}[section]
\newtheorem{definition}{Definition}[section]
\newtheorem{proposition}{Proposition}[section]
\begin{document}
\title{TRANSITION PROBABILITIES BETWEEN\\ QUASIFREE STATES}

\author{H. Scutaru}
\address{Department of Theoretical Physics, Institute of Atomic Physics\\
Bucharest-Magurele, POB MG-6, Romania\\
e-mail: scutaru@theor1.ifa.ro}
\maketitle

\begin{abstract}
We obtain a general formula for the transition probabilities
between any state of the $C^*$ algebra of the  canonical
commutation relations (CCR-algebra) and a squeezed quasifree 
state (Theorem III.1). Applications of this formula are made for the
case of multimode thermal squeezed states of quantum
optics using a general canonical decomposition of
the correlation matrix valid for any quasifree state.
In the particular case of a one mode CCR-algebra we
show that the transition probability between two quasifree
squeezed states is a decreasing function of the geodesic 
distance between the points of the upper half plane 
representing these states. In the special case of
the purification map it is shown that the
 transition probability between
the state of the enlarged system and the product state of real
and fictitious subsystems can be a measure for the entanglement. 
\end{abstract}

\section{Introduction}

   The notion of the quasifree state has appeared and was developed 
in the
framework of the $C^*$-algebra approach to the canonical commutation 
relations (CCR) \cite{kas,man1,man2,van,fan,hol1,hol2,hol3,hol4,scu1,scu2}. 
The quasifree states are 
the natural ground 
states for the Hamiltonians which are at most quadratic in the 
bosonic creation
and annihilation operators. The essential property of the quasifree
states is the fact that all their correlations are expressible in
terms of the one and two-point functions.

The advantage of the $C^*$-algebra approach to the theory of 
coherent and squeezed states (thermal or not) follows from the fact
that the results are obtained in a representation independent form.
This representation independence allows to elude the hard noncommutative
calculation. But a price must be paid for this. A simple result
in usual quantum mechanics is obtained, in this case, with the help of 
slightly difficult results concerning the transition probabilities
between the states of $C^*$-algebras \cite{alb}.

We shall illustrate this assertion by giving a representation
independent formula for the transition probability between a
squeezed state and an arbitrary state of the $C^*$-algebra of CCRs
(Theorem III.1).

Many interesting states of the radiation field in quantum optics
are quasifree states. The factorization of the correlations
in coherent states was discovered by Glauber as the essential
property of these states of coherent light. The squeezed states
are also quasifree states. The most general case is that of
multimodal thermal displaced and squeezed states.

The main result of the paper, given in Section III, is the formula
for the transition probabilities between any state of the $C^{*}$-
algebra of the canonical commutation relations and any squeezed
quasifree pure state (Theorem III.1). This result is obtained using the fact
that any quasifree squeezed pure state can be obtained
from a coherent state by a Bogoliubov automorphism and the fact that
for any coherent state there is an element of the CCR-algebra
which is a minimal orthoprojection \cite{van}. Then a general result of
P. M. Alberti and V. Heinemann \cite{alb} is applicable.

A new approach to the theory of multimode squeezed thermal
states is described in Section IV, and is based on the
combination of the Williamson theorem \cite{fol} which gives
the most general structure of a positive definite matrix
 with a theorem
of Balian, De Dominicis and Itzykson \cite{bal} which gives the
most general structure of a symplectic matrix.
In this way a general natural parametrization of the correlation
matrix of a quasifree state is obtained. This parametrization
points out the squeezing and the orthogonal symplectic
transformations. The orthogonal transformations mix the
different modes.

The geometric  interpretation, given in Section V, is valid
only for the one mode case and shows that the transition
probability between two pure squeezed states decreases with the
increase of the geodesic distance between the corresponding points
of the Poincar\'e upper half plane.

In Section VI the relation between the Wigner function of a
quasifree state of a system and that of a subsystem is given in
most general form. A first application of this relation is given
in Section VII where the correlation matrix of a pure quasifree state
of an enlarged system ( which reduces to the mixed state of the 
orginal system ) is obtained in an elementary way \cite{van}.

In Section VIII it is shown that the transition probability between
the above pure state obtained for the enlarged system and the
mixed state obtained as the direct product of the state
of the original subsystem with the state of the fictitious
subsystem can be a measure of the entanglement of these two
subsystems reflected in the pure state of the total system.

Finally, in Section IX, it is shown how the correlations between the
original subsystem and the fictitious one are created
by the Bogoliubov transformations which map the Fock
state into the above pure state obtained by the purification
procedure.

\section{Quasifree states}

The phase space considered in the present paper is a finite
dimensional symplectic space $ (E,\sigma) $. This is a real
vector space endowed with a real, bilinear, antisymmetric 
form $ \sigma(.,.) $ which gives the symplectic structure on $E$.
Then $E$ is of even real dimension $2n$ and there exist in $E$
symplectic bases of vectors $\{e_j, f_j\}_{j=1,...,n}$ , i.e.
reference systems such that $\sigma(e_j,e_k) = \sigma(f_j,f_k)=0$
and $\sigma(e_j,f_k) = - \sigma(f_k,e_j) = \delta_{jk}$, $j,k=1,...,n$.
The coordinates $(\xi^j,\eta^j)$ of a vector $u\in E$ in a
symplectic basis $( u= \sum_{j=1}^n(\xi^je_{j}+\eta^jf_{j}) )$ are
called symplectic coordinates. The measure $dm(u) = \prod_{j=1}^n
d\xi^jd\eta^j$ is the same for all symplectic coordinates and
is called the Liouville measure on $(E,\sigma)$ \cite{kas}. 
There is a one-to-one
correspondence between the symplectic bases and the linear
operators $J$ on $E$ defined by $Je_k=-f_k $ and $Jf_k=e_k$,
$k=1,...,n$. The essential properties of these operators are:
$\sigma(Ju,u)\geq 0$, $ \sigma(Ju,v)+\sigma(u,Jv)=0$ ($u,v \in E$ and
$J^2=-I$, $I$ denotes the identity operator on $E$).Such operators
are called complex structures \cite{kas}.
In the following
we shall use the matricial notations with $u\in E$ as column vectors.
Then $\sigma(u,v)=u^TJv$ and the scalar product is given by 
$\sigma(Ju,v)=
u^Tv, u,v\in E$. A linear operator $S$ on $E$ is called a
symplectic operator if $S^TJS=J$. When $S$ is a symplectic operator
then $S^T$ and $S^{-1}$ are also symplectic operators. The
group of all symplectic operators $Sp(E,\sigma)$ is called the
symplectic group of $(E,\sigma)$. The Lie algebra of $Sp(E,\sigma)$
is denoted by $sp(E,\sigma)$ and its elements are operators $R$
on $E$ with the property: $(JR)^T=JR$.
If $J$ and $K$ are two complex
structures, there exists a symplectic transformation $S$
such that $J = S^{-1}KS$. 
\begin{definition}\cite{kas,man2}
The $C^*$-algebra {\bf A} of the canonical commutation relations (CCR)
is obtained 
by the completion of the *-algebra 
Span\{ $\delta_u:u\in E$\}, the elements of which satisfy the Weyl 
relations:
\begin{equation} 
 \delta_u \delta_v = exp(-i\sigma(u,v)/2) \delta_{u+v} , 
 \delta_u^* = \delta_{-u} , u,v\in E    
\end{equation}
\end{definition}  
   A state on {\bf A} is a positive linear functional $\omega$
normalized by $\omega(\delta_0) = 1$
\cite{kas,man1,man2,van,fan,hol1,hol2,hol3,hol4,scu1,scu2}. 
The complex valued 
function
on the phase space $\omega(\delta_u)$ is called the characteristic
or the generating function of the state $\omega$. A continuous 
complex valued function $f$ on $E$ , with $f(0)=1$, is the
characteristic function of a state on {\bf A} if and only if:   
\begin{equation}
 \sum_{j,k=1}^{N} \bar a_{j}a_{k} exp(i\sigma(u_{j},u_{k})/2) 
f(u_{k}-u_{j}) \geq 0
\end{equation} 
for all $a_{k} \in {\bf C}$, $u_{k} \in E$, $k=1,...,N$, and 
$N\in {\bf N}$.
 
For any $v \in E$ an automorphism $\tau_{v}$ of $\bf A$ can be 
defined by:
\begin{equation}  
 \tau_{v}(x) = \delta_{v}^{*} x \delta_{v}    
\end{equation}
A Bogoliubov transformation of {\bf A} is a *-automorphism of {\bf A}
defined for any $S\in Sp(E,\sigma)$ by
\begin{equation} 
\alpha_{S}(\delta_{u}) = \delta_{Su} , u \in E. 
\end{equation} 
The quasi-free automorphisms of {\bf A} are those of the form
$\tau_{v} \circ \alpha_{S} $ where $v \in E$ and $S \in Sp(E,\sigma)$.

   Let A be a symmetric $(A^{T}=A)$ and positive $(u^{T}Au \geq 0$ for all 
$u \in E)$ matrix. 
\begin{definition}\cite{kas,man1,man2,van,fan,hol1,hol2,hol3,hol4,scu1,scu2}
A quasi-free state $\omega_{(A,v)}$ on $\bf A$ is 
a state whose characteristic function is given by the formula:
\begin{equation} 
\omega_{(A,v)}(\delta_{u}) = exp(-u^{T}Au/4 + iu^{T}v) 
\end{equation}
with the restriction:
\begin{equation}
 -JAJ \geq A^{-1} 
\end{equation}
\end{definition}
Evidently, we have:
\bea 
\omega_{(A,v)} \circ \alpha_{S} = \omega_{(S^{T}AS,S^{T}v)} 
\eea
and $\omega_{(A,v)} = \omega_{(A,0)} \circ \tau_{v} $.
The state $\omega_{(A,v)} $ is pure iff $\omega_{(A,0)}$ is pure and this
is the case iff $-(JA)^2 = I$ ( this is equivalent with $JA \in Sp(E,\sigma)$ 
) and, in particular, when $A = I$ i.e when the state is a coherent state.
The pure squeezed states are the quasi-free states of the form 
$\omega_{(I,v)} \circ \alpha_{S} = \omega_{(S^{T}S, S^{T}v)}$.

\section{Transition probabilities}

   The transition probability $P(\omega_{1};\omega_{2})$ between
two arbitrary states $\omega_{1}$ and $\omega_{2}$ on ${\bf A}$
can be defined in many ways \cite{bur,uhl,ara,alb}. 
\begin{theorem}
If the state $\omega_{1}$ is a squeezed state the transition 
probability to an arbitrary state $\omega_{2}$ can be computed by the 
following representation independent formula:
\bea 
P(\omega_{1};\omega_{2}) = (2 \pi )^{-n} \int_{E} \omega_{1}(\delta_{-u})
\omega_{2}(\delta_{u}) dm(u) 
\eea 
where $dm(u)$ is the Liouville measure defined above. 
\end{theorem}
A proof of this
result can be obtained by combining the construction of a projection as
in Lemma 3 from \cite{van} with the assertion (vi) of Sec. II from \cite{alb}.

\begin{lemma}
For any coherent state $\omega_{(I,v)}$ we can define an
element  $p_{v} \in {\bf A}$ such that it is a projector (i.e.  
$p_{v}^{2} = p_{v}$ and $p_{v}^{*} = p_{v}$)
and has the property
\bea 
p_{v}xp_{v} = \omega_{(I,v)}(x)p_{v} 
\eea
for any $x \in {\bf A}$.
\end{lemma}
{\bf Proof}. We take
\bea 
p_{v} = (2 \pi)^{-n} \int_{E} \omega_{(I,v)}(\delta_{-u})
\delta_{u}dm(u) 
\eea  
The fact that $p_{v}$ is a selfadjoint
element of ${\bf A}$ is evident. The fact that it is an idempotent
element of ${\bf A}$ results from
\begin{eqnarray} 
&&
\nonumber
p_{v}^{2} = (2 \pi)^{-2n} \int_{E} \int_{E}
\omega_{(I,v)}(\delta_{-u}) \omega_(\delta_{-w}) \delta_{u} \delta_{w}
dm(u)dm(w) =\\ 
&&
\nonumber
(2\pi)^{-2n} \int_{E} 
\omega_{(I,z)}(\delta_{-z}) I(z) 
dm(z)\\ 
\end{eqnarray}
where
$I(z) = \int_{E} {\omega_{(I,v)}(\delta_{-u}) 
\omega_{(I,v)}(\delta_{-z+u}) \over \omega_{(I,v)}(\delta_{-z})}
exp(-{i \over 2} \sigma(u,z)) dm(u)  $.
Thus it is sufficient to prove that $I(z) = (2\pi)^{n}$ and this
follows by direct computations.
In this way we have associated to any coherent state $\omega_{(I,v)}$
a projector from ${\bf A}$. 
In order to prove the last assertion it is sufficient to prove that:
\begin{eqnarray} 
&&
\nonumber
\int_{E} \omega_{(I,v)}(\delta_{-w})\omega_{(I,v)}(\delta
_{-y+w+u}) \exp\{ -{1\over 2}i(\sigma(w,u)+\sigma(w,y))\}
\\ 
&&
\nonumber
dm(w)=(2\pi)^{n}\omega_{(I,v)}(\delta_{u})
\omega_{(I,v)}(\delta_{-y}) \exp {1\over 2}i \sigma(u,y) \\
\end{eqnarray}
As it results from \cite{kas,man1,man2} the property (3.5) 
is valid for any $x \in {\bf A}$ and (3.2) follows.                                                            

Then we can apply the following result of P.M. Alberti and 
\newline V.Heinemann 
given in the Section II of the paper \cite{alb} at the point (vi):
\begin{lemma}
Let ${\sl A}$ a $C^{*}$-algebra and ${p \in {\sl A}}$ be a
minimal orthoprojection, i.e., $p{\sl A}p = {\bf C}p$.
Let $pxp = \nu(x)p$, with $\nu(x) {\in {\bf C}}$. Then
the map $\nu : x \rightarrow {\nu(x)\in {\bf C}}$ defines
a state ${\nu \in S({\sl A}) }$ (the set of states on 
{\sl A}) with $\nu(p) = 1$ and we have 
\bea 
P(\omega;\nu) = \omega(p) 
\eea
for any state ${\omega \in S({\sl A})}$.
\end{lemma}
{\bf Proof of the Theorem}.
We must verify the fact that
\bea 
\omega_{(I,v)}(p_{v}) = 1. 
\eea
We have 
\bea 
\omega_{(I,v)}(p_{v}) = (2\pi)^{-n}\int_{E}exp(-{1\over2}w^{T}w)
dm(x) = 1  
\eea

Thus we have proved that for any state $ \omega $ and  coherent
state $ \omega_{(I,v)} $ the transition probability is given by :
\bea 
P(\omega;\omega_{(I,v)}) = \omega(p_{v}) = (2\pi)^{-n}\int_{E}
\omega_{(I,v)}(\delta_{-v})\omega(\delta_{v})dm(v)
\eea

\begin{corrolary} 
The general formula for the transition probability
between any quasifree state and a pure quasifree
state is given by:
\bea  
P(\omega_{(A,v)}, \omega_{(B,w)}) = (det({A+B \over 2}))^
{-{1 \over 2}}exp(-(w-v)^{T}(A+B)^{-1}(w-v))  
\eea
\end{corrolary}
{\bf Proof}.
It is well known \cite{bur,uhl,ara,alb}
that for any automorphism $\alpha$ of ${\bf A}$ we have :
\bea 
P(\omega_{1} \circ \alpha;\omega_{2} \circ \alpha) =
P(\omega_{1};\omega_{2}) 
\eea
When the state $\omega$ is a quasi-free state $\omega_{A}$ we
have
\bea 
P(\omega_{A};\omega_{(I,v)}) = det({I+A\over2})^{-1/2}
 exp(-v^{T}(I+A)^{-1}v) 
\eea

The transition probability between two squeezed states can be
also given if we take into account the fact that any squeezed
state is obtained by a Bogoliubov automorophism from a coherent
state and the invariance of the transition probabilities with
respect to automorphisms.
\newline
{\bf Remark}. When both states are squeezed pure states we have:
\bea 
P(\omega_{(A,v)};\omega_{(B,w)}) = det({I+S^{T}S\over2})^{-1/2}
exp(-(w-v)^{T}(A+B)^{-1}(w-v)) 
\eea
where $S$ is the symplectic matrix such that $A = S^{T}BS$.

\section{Applications in quantum optics}

\begin{theorem}
The most general form of a correlation matrix $A$ is given by:  
\bea 
A = O^{'T}\left(\matrix{M&0\cr 0&M^{-1}\cr}\right)O^{T}        
\left(\matrix{D&0\cr 0&D\cr}\right)O\left(\matrix{M&0\cr 0&M^{-1} 
\cr}\right)O^{'} 
\eea
where $O$ and $O^{'}$ are symplectic and orthogonal $(O^{T}O=I)$
operators and M is a diagonal $n\times n$ matrix.
\end{theorem}
{\bf Proof}.
The proof results by combining two well known results.
Because the correlation matrix A is positive definite it
follows \cite{fol} 
that there exists $S \in Sp(E,\sigma)$ such that
\bea 
A = S^{T}\left(\matrix{D&0\cr 0&D\cr}\right)S 
\eea
where $D$ is a diagonal $n\times n$ matrix.

   The most general real symplectic transformation $S\in Sp(E,\sigma)$
has \cite{bal}
the following structure:
\bea 
S = O\left(\matrix{M&0\cr 0&M^{-1}\cr}\right)O^{'} 
\eea
where $O$ and $O^{'}$ are symplectic and orthogonal $(O^{T}O=I)$
operators and M is a diagonal $n\times n$ matrix.
\newline {\bf Examples}. 
Various particular kinds of such matrices are obtained taking $O$,
$O^{'}$, $D$ or $M$ to be equal or proportional to the corresponding 
identity operator.
A pure squeezed state is obtained when $D=I$. If this condition
is not satisfied, the state is a mixed state called thermal squeezed
state \cite{eza}. When $M=I$ there is no squeezing and the correspondig states
are pure coherent states or thermal coherent states. All these states
have correlations between the different modes produced by the
orthogonal symplectic operators $O$ and $O^{'}$. Combining Eq.(3.2)
and Eq.(3.12) we obtain that:  
\bea  
P(\omega_{A} \circ \alpha_{S};\omega_{(I,0)}) = 
P(\omega_{A};\omega_{(I,0)})
\eea 
iff $S^{T}AS=A$ or iff $S^{T}S=I$ i.e. iff S is orthogonal.

The first situation appears for example in the case of a parametric
amplifier \cite{hua}
and this property explains the fact that the
nonentanglement of states is preserved by such a transformation.
Hence the parametric amplifier is a transition probability 
preserving device.
The second situation appears if there is no squeezing i.e. $M=I$. 
As it was shown in \cite{hol2}
for any quasifree state $\omega_{A}$ we have:
\bea \omega_{A} =(\omega_{d_{1}}\otimes \omega_{d_{2}}\otimes...
\otimes \omega_{d_{n}}) \circ \alpha_{S} \eea 
where $d_{1},...,d_{n}$ are the diagonal elements of the matrix $D$,
and where $\omega_{d_{i}}$, $i=1,...,n$ are the one mode quasi-free
states defined on the phase spaces $E_{i}$ generated by the vectors
$\{e_{i},f_{i}\}$ (on which the operator $A$ acts by $Ae_{i}=d_{i}e_{i}$
and $Af_{i}=d_{i}f_{i}$) by the formula:
\bea 
\omega_{d_{i}}(\delta_{\xi^{i}e_{i}+\eta^{i}f_{i}}) =  
exp(-d_{i}((\xi^{i})^2+(\eta^{i})^2)/4) 
\eea 
with $d_{i} \geq 1$ as it follows from Eq.(2.6).
It is clear that
\bea 
P(\omega_{d_{1}}\otimes \omega_{d_{2}}\otimes...
\otimes \omega_{d_{n}};
\omega_{(I,0)}) = \prod_{i=1}^n  P(\omega_{d_{i}};\omega_{(I,0)}) 
\eea
Combining Eqs.(4.5) and (4.7) we obtain that the equality
\bea 
P(\omega_{A};\omega_{I}) = \prod_{i}^n
P(\omega_{d_{i}};\omega_{I}) 
\eea
is valid only for nonsqueezed states (i.e. when$M=I$).

For pure squeezed states $(D=I)$ we still have such a decomposition
of the transition probability as a product of transition probabilities
for the individual modes. Moreover, there is another class of quasi-free
states defined by the condition $O=I$
in Eq. (4.3), and for which such a decomposition is valid.
The formula which describes these two situations is: 
\bea P(\omega_{A};\omega_{I}) = \prod_{i}^n {2 \over
\sqrt{(m_{i}^2+d_{i})(m_{i}^{-2}+d_{i})}} \eea
The thermal squeezed states considered in [18] 
are obtained when
$M=mI$ and $D=dI$. In this case the correlation matrix $A$ takes the
following form:
\bea 
A = O^{'T} \left(\matrix{m^{2}dI&0\cr 0&m^{-2}dI\cr} \right)
O{'} 
\eea
and
\bea 
P(\omega_{A};\omega_{I}) = \left({2 \over \sqrt{
(m^2+d)(m^{-2}+d)}} \right)^n 
\eea
     
We shall take into account the fact that the most general form of 
an orthogonal symplectic matrix is \cite{bal,fol}: 
\bea 
O^{'} = \left(\matrix{X&Y\cr -Y&X\cr} \right) 
\eea
where $X$ and $Y$ are $n \times n$ matrices which satisfy the 
conditions: $X^{T}X +Y^{T}Y=I$ and $X^{T}Y=Y^{T}X$.
 
When $Y=0$ the correspondig quasifree state is nonentangled
and remains nonentangled in all symplectic frames obtained
by such Bogoliubov transformations (i.e. orthogonal with $Y=0$) 
\cite{eza}.
The ideal (lossless) beam splitter effects a transformation of
such kind and the nonentanglement is preserved only
when all modes are with the same temperature and equally
squeezed \cite{hua}. It follows from the above discussion that the ideal beam
splitter is a transition probability preserving device.

\section{A geometric interpretation}

The squeezed states are in a one-to-one correspondence with the 
elements of 
\newline ${\bf R} = sp(E,\sigma) \bigcap Sp(E,\sigma)$. 
This correspondence
is covariant with respect to the adjoint action of 
the symplectic
group $Sp(E,\sigma)$ on ${\bf R}$ and the action of this group
on the squeezed states given by the corresponding 
Bogoliubov authomorphisms. Because the symplectic group acts
transitively on ${\bf R}$ it follows that this is a coadjoint                 
orbit of $Sp(E,\sigma)$. In fact, this is the Hermitian symmetric
space $Sp(E,\sigma)/U(n)$ , where $U(n)$ is the subgroup of 
$Sp(E,\sigma)$ the elements of which are also orthogonal.
Hence the squeezed states are labeled by the elements of this
Hermitian symmetric space.
In the one mode case $(n=1)$ the Hermitian symmetric space
$Sp(2,R)/U(1)$ is the Poincar\'e upper half plane ${\bf H}$ \cite{lang}.
The group $Sp(2,R)$ acts on ${\bf H}$ in the usual way:
\bea 
\gamma : z \longrightarrow {az+b\over cz+d} 
\eea
where $z=x+\sqrt{-1}y$ and where
\bea 
\gamma = \left(\matrix{a&b\cr c&d\cr} \right) 
\eea
with $det(\gamma) = ad-bc=1$.
The function 
\bea 
u(z,z^{'}) ={ \vert z - z^{'} \vert ^2\over 4yy^{'}} 
\eea
is invariant to the action of the group $Sp(2,R)$ :
\bea 
u(\gamma z,\gamma z^{'}) = u(z,z^{'}). 
\eea
The Poincar\'e metrics on ${\bf H}$ is also invariant and is
given by :
\bea 
ds^2 ={dx^2 + dy^2 \over y^2} = {dz d \bar z \over y^2} 
\eea
The corresponding distance function $\rho(z,z^{'})$ is equal
with the lenght of the geodesic between $z$ and $z^{'}$.
\begin{proposition}
For any pair of squeezed states 
$\omega_{z}$ and $\omega_{z^{'}}$:
\bea 
P(\omega_{z};\omega_{z^{'}}) = {1 \over ch(s/2)} 
\eea
where $s$ is the geodesic distance between $z$ and $z^{'}$. 
\end{proposition}
{\bf Proof}.
For any two points $z,z^{'} \in {\bf H}$ there exists
$\gamma \in Sp(2,R)$ such that $\gamma z = \sqrt{-1}$ and
$\gamma z^{'} = \sqrt{-1}y_{0}$ for a real $y_{0} \geq 1$.
Evidently $\rho(z,z^{'}) = \rho(\sqrt{-1},\sqrt{-1}y_{0})$.
The geodesic between $\sqrt{-1}$ and $\sqrt{-1}y_{0}$ is the
vertical straight line connecting these two points \cite{lang}. Then
\bea 
\rho(\sqrt{-1},\sqrt{-1}y_{0}) = \int_{1}^{t}{dy \over y} =
ln(t) = s 
\eea
and $1+u(\sqrt{-1},\sqrt{-1}t) = ch^2(s/2)$. 
From $\gamma \sqrt{-1} = {a\sqrt{-1}+b \over c \sqrt{-1}+d} =
x+ \sqrt{-1}y$ we obtain $x={bd+ac \over c^2+d^2}$ and
$y={ad-bc \over c^2+d^2}$ and 
\bea 
1 + u(z,\sqrt{-1}) = {a^2+b^2+c^2+d^2+2 \over 4} 
\eea
The correspondence between the points of the Poincar\'e half
plane {\bf H} and the squeezed states is given by
\bea 
\sqrt{-1} \longleftrightarrow J= \left(\matrix {0&1\cr -1&0\cr
}\right) 
\eea
and $z \longleftrightarrow J \gamma$.
Then
\bea  
P(\omega_{z},\omega_{\sqrt{-1}}) = det({1+\gamma^{T}\gamma
\over 2})^{-1/2} = {1 \over ch(s/2)} 
\eea
Evidently, this result is valid for any pair of squeezed states. 

\section{The reduction of the Wigner functions to the subsystems}

In the following we shall use the fact that the Wigner
functions of the states on the algebra of the bosonic
commutation relations give a complete description of
these states.
\begin{definition}
The Wigner function of the state $\omega$ is defined as
the Fourier symplectic transform of the corresponding
characteristic function
$\omega(\delta_{u})$: 
\bea 
W_{\omega}(u)\;=\;(2\pi)^{-2n}\int_{E}\exp[i\sigma(u,v)]
\omega(\delta_{v})dm(v)
\eea
\end{definition}
\begin{proposition} 
The Wigner function of a quasi-free state is given by  
\bea 
W_{\omega}(u)\;=\;\pi^{-n}(detG)^{1\over2}\exp(-u^TGu)
\eea
where $G=-JA^{-1}J$ is a  $2n \times 2n$
real, symmetric and
positive definite matrix which satisfies a
restriction which is equivalent with the restriction (2.6):
\bea
-(JG)^2 \leq I 
\eea
\end{proposition}
{\bf Remarks}.
We have also $JG\in sp(E,\sigma)$. If $\omega$ is a
pure quasi-free state then $G=A$ \c and $detG=detA=1$
(because $G$ and $A$ belong to $Sp(E,\sigma)$). 
\begin{corrolary}
The most general form of the matrix $G$ which defines the Wigner function
of a quasi-free state is the following:
\bea 
G = O^{'T}\left(\matrix{M&0\cr 0&M^{-1}\cr}\right)O^{T}        
\left(\matrix{D^{-1}&0\cr 0&D^{-1}\cr}\right)O\left(\matrix{M&0\cr 0&M^{-1} 
\cr}\right)O^{'} 
\eea
where $O$ and $O^{'}$ are symplectic and orthogonal $(O^{T}O=I)$
matrices and M is a diagonal $n\times n$ matrix.
\end{corrolary}
{\bf Proof}
Since $S^TJS=(S^{-1})^TJS^{-1}=SJS^T=J$ for any
symplectic matrix $S$ it follows that
\bea 
G= -JA^{-1}J = -JS^{-1}\left(\matrix{D&0\cr0&D\cr}\right)
(S^T)^{-1}J= S^T\left(\matrix{D^{-1}&0\cr0&D^{-1}\cr}\right)S
\eea
  The most general real symplectic transformation $S\in Sp(E,\sigma)$
has [17] 
the following structure :
\bea 
S = O\left(\matrix{M&0\cr 0&M^{-1}\cr}\right)O^{'} 
\eea
where $O$ and $O^{'}$ are symplectic and orthogonal $(O^{T}O=I)$
operators and M is a diagonal $n\times n$ matrix.
\newline As a consequence we have that $detG=(detD)^{-2}$.

Let us supose that the state of the quantum system ${\cal S}$ 
is described by the Wigner function $W(u)$ defined
on the phase space $(E, \omega)$ which can be considered
as being composed from two subsystems ${\cal S}_{1}$ and ${\cal S}_{2}$ 
with the phase spaces $(E_{1}, \omega_{1})$ and $(E_{2}, \omega_{2})$
respectively, where $E = E_{1} \bigoplus E_{2}$ and  
$\omega_{1}$ and $ \omega_{2}$ are the restrictions of
the state $\omega$ to $E_{1}$ and to $E_{2}$ respectively.
In this case the Wigner function 
$W_{\omega}(u)=\pi^{-n}(detG)^{1\over2}\exp(-u^TGu)$
can be writen as
\bea 
W(x,y) = \pi^{-n}
\left(det\left(\matrix{A&B\cr B^T&C\cr}\right)\right)  
^{1\over2} exp-(x^T~y^T)\left(\matrix{A&B\cr B^T &C\cr}\right)
\left(\matrix{x \cr y}\right) 
\eea
\begin{theorem}
If $detC \neq 0$ then the Wigner function of the subsystem ${\cal
S}_{1}$ is given by
\bea W_{1}(x) = \pi^{-{n \over 2}}(det(A-BC^{-1}B^T) )^{{1 \over 2}} 
exp\left[-x^T(A-BC^{-1}B^T)x\right] \eea
and if $detA \neq 0$ the Wigner function of the susystem
${\cal S}_{2}$ is given by
\bea 
W_{2}(y) = \pi^{-{n \over 2}}(det(C-B^TA^{-1}B))^{{1 \over 2}} 
exp\left[-y^T(C-B^TA^{-1}B)y\right] 
\eea
\end{theorem}
{\bf Proof}. The Wigner function $W_{1}(x)$ of the subsystem ${\cal S}_{1}$ is 
defined by the integration on the  coordinates of the subsystem
${\cal S}_{2}$
\bea 
W_{1}(x) = \int_{E_{2}} W(x,y) dm(y) 
\eea
If $detC \neq 0$ then
\bea 
W_{1}(x) = \pi^{-{n \over 2}}(det(A-BC^{-1}B^T) )^{{1 \over 2}} 
exp\left[-x^T(A-BC^{-1}B^T)x\right] 
\eea
We have used Schur's formula
\bea 
det \left(\matrix{A&B\cr B^T &C\cr}\right)=
detC det(A-BC^{-1}B^T) 
\eea
Analogously, if $detA \neq 0$ then
\bea 
W_{2}(y) = \pi^{-{n \over 2}}(det(C-B^TA^{-1}B))^{{1 \over 2}} 
exp\left[-y^T(C-B^TA^{-1}B)y\right] 
\eea
Also we have used the fact that
\bea 
det \left(\matrix{A&B\cr B^T &C\cr}\right)=
detA det(C-B^TA^{-1}B) 
\eea

\section{The purification of a mixed state}
\begin{theorem}
For any quasifree mixed state $\omega_{A}$ on the $C^*$-algebra 
of commutation relations defined on the phase space
$E$ there exists a pure quasifree state $\widetilde\omega_{A}$ 
on the $C^{*}$-algebra of commutation relations defined on the
phase space $E \oplus E$ endowed with the symplectic structure
$\left(\matrix{J&0\cr0&-J\cr}\right)$. 
This pure quasi-free state 
$\widetilde\omega_{A}$ is defined by the correlation matrix ${\widetilde A}$
(which is a $4n\times4n$ real, symmetric and positive definite matrix) 
\bea 
{\widetilde A} = \left(\matrix{A & A\sqrt{I+(JA)^{-2}}  \cr 
 A\sqrt{I+(JA)^{-2}}  & A \cr}\right)
\eea
\end{theorem}
{\bf Proof}.
This pure quasi-free state 
$\widetilde\omega_{A}$ is defined by the correlation matrix ${\widetilde A}$
(which is a $4n\times4n$ real, symmetric and positive definite matrix) 
\bea 
{\widetilde A} = \left(\matrix{U & V \cr  V & U \cr}\right)
\eea
which satisfies the restriction
\bea
-\left(\matrix{
\left(\matrix{J & 0 \cr 0 & -J \cr} \right) \left(\matrix{
U & V \cr V & U \cr} \right)} \right)^{2} = 
\left(\matrix{I & 0 \cr 0 & I}\right)
\eea       
i.e. 
\begin{equation}
(JU)^2-(JV)^2= -I,~~~  JUJV=JVJU
\end{equation}
The map which associates a pure quasifree state 
$\widetilde\omega_{A}$ 
to any mixed quasi-free state $\omega_{A}$ is called the
purification map. 
Because $\widetilde\omega_{A}$ is a pure quasi-free state
the corresponding Wigner function is defined by the
same matrix:
\bea 
\tilde G= \left(\matrix{U & V \cr V & U \cr}\right)
\eea
We remark that the original system can be considered as
the reduction of the extended one. Then one must have the relation 
\begin{equation}
\int_{E} W_{\widetilde \omega_{A}}((u,\widetilde u))dm(\widetilde u)
=W_{\omega_{A}}(u)
\end{equation} 
Hence we can use the result of the
preceeding section i.e. we have 
\begin{equation}
G_{1}=U-VU^{-1}V=G=-JA^{-1}J
\end{equation}
From the Schur formula it follows that
\bea 
det \tilde G = detU det(U- VU^{-1}V)=detA detA^{-1}
=1
\eea
From the equations (7.3) and (7.6) we have 
\newline $JG=JU-(JV)^2(JU)^{-1}$.
Hence $JGJU=(JU)^2-(JV)^2=-I$, from which it follows that
$U=A$ and $(JV)^2 = I + (JA)^2$. It is easy to prove that
$V= A\sqrt{I+(JA)^{-2}}$ is a solution of the last equation.
\newline {\bf Remarks}. The map defined by (7.6) which acts on 
the Wigner functions is the inverse map of the purification map. 
From the formula:
\begin{eqnarray} 
W_{\widetilde\omega_{A}}((u,\widetilde u)) = \pi^{-2n}
\exp\{-u^TAu-\widetilde u^TA\widetilde u -
V^T\widetilde u
-\widetilde u^TVu)\}
\end{eqnarray}
it follows that
the correlations between the variables $u$ of the real system
and the variables $\widetilde u$ of the fictitious one are given
by the terms $(A\sqrt{I+(JA)^{-2}}u)^T\widetilde u + \widetilde u^T(A\sqrt{I+
JA)^{-2}} u)$ which appear at the exponent. 

In the general case
\bea 
A = S^T \left(\matrix{D&0\cr0&D\cr}\right)S 
\eea
and $JS^T = S^{-1}J$ and $J\left(\matrix{D&0\cr0&D\cr}\right)
=\left(\matrix{D&0\cr0&D\cr}\right)J$.
From these relations it follows that
\begin{eqnarray}
&&
\nonumber
(JA)^{-2} = A^{-1}JA^{-1}J= \\
&&
\nonumber
S^{-1}
\left(\matrix{D^{-1}&0\cr0&D^{-1}\cr}\right)(S^T)^{-1}J
 S^{-1}
\left(\matrix{D^{-1}&0\cr0&D^{-1}\cr}\right)(S^T)^{-1}J = \\
&&
\nonumber
- S^{-1}
\left(\matrix{D^{-2}&0\cr0&D^{-2}\cr}\right)S\\
\nonumber
\end{eqnarray}
Hence
\begin{eqnarray}
&&
\nonumber
I+(JA)^{-2} = I - S^{-1}
\left(\matrix{D^{-2}&0\cr0&D^{-2}\cr}\right)S=\\
&&
\nonumber
S^{-1}
\left(\matrix{I-D^{-2}&0\cr0&I-D^{-2}\cr}\right)S\\
\nonumber
\end{eqnarray}
and
\begin{eqnarray}
A\sqrt{I+(JA)^{-2}}=
S^{T}\left(\matrix{\sqrt{D^{2}-I}&0\cr0&\sqrt{D^{2}-I}\cr}\right)S
\end{eqnarray}

\section{A measure of the entanglement}

A characteristic feature of quantum mechanics is the
presence of the correlations between the subsystems of
a quantum systems described by a pure state. In this
situation the states of the subsystems are mixed.
Any quantum system can be considered to be a subsystem
of a larger quantum system which is in a pure state.
The enlarged system can contain real or fictitious additional
subsystems. The appearance of the quantum correlations
between these auxiliary subsystems and the initial one
is the price which must be payed for the purity of
the state of the whole system.

For any mixed quasi-free state on a n-mode  
bosonic system there exists a standard procedure which
defines a compound 2n-mode boson\-ic system in a pure
quasifree state such that the restriction of this
state to the initial subsystem is the initial mixed quasi-free
state. This procedure is called the purification operation.

The corresponding compound system has two subsystems
which are identical with the initial one and which
are strongly correlated. These correlations are
explicitely defined by the purification operation.

In Section III we have obtained the following 
general formula for the transition probability
between any quasifree state and a pure quasifree
state:

\bea  
P(\omega_{A}, \omega_{B}) = (det({A+B \over 2}))^
{-{1 \over 2}}  
\eea

Because the state obtained by purification is a pure state
we can apply this formula in the case in which the
mixed quasifree state is that given by the direct product
of the mixed states of the real and fictitious systems.
The last one is a mixed quasifree state with the
correlation matrix given by:
\bea 
\left(\matrix{A&0\cr0&A\cr}\right) 
\eea
Hence we must find 
\bea 
det\left(\matrix{A & {A \over 2} \sqrt{I+(JA)^{-2}} 
\cr 
{A \over 2}\sqrt{I+(JA)^{-2}}
& A \cr}\right)\eea 
One obtains
\bea  
P(\omega_{A \bigoplus A}, \tilde \omega_{A}) 
= (det({3 \over 4}D^2 + {1 \over 4}I))^
{-1}  
\eea
This transition probability can be considered as a measure
of the correlation between the real and fictitious subsystems.
One can remark that this transition probability is a
decreasing function of the temperature of each mode i.e. the
correlation is increasing with the temperature.

\section{Correlations that are given by the Bogoliubov automorphisms}

Another useful result of the theory of bosonic commutation
relations which can be applied to the pure state which
appears from the purification operation is the fact that
all pure quasifree states can be obtained from the Fock state
by a Bogoliubov transformation. In other words the pure
quasifree states are squeezed states in a generalized
meaning.
Hence the strong correlations between the real modes
and the fictitious ones are produced by the action of
the Bogoliubov automorphisms which are of two kinds:
squeezing transformations of each mode and transformations
which mix different modes.

In the following we shall disscuss the case of pure quasi-free
states in detail. 
If $\omega_{A}$  is a pure quasi-free state then $-(JA)^{2}=I$.
From the Williamson theorem \cite{fol} for any positive definite symmetric
matrix $A$ there exists an element $S  \in Sp(E,\sigma)$ 
and a diagonal matrix with positive entries $D$ and with the 
property $DJ=JD$ and such that $A=S^TDS$. Then from the
restriction  $-(JA)^{2}=I$ one obtains  $D^{2}=I$
i.e. $D=I$. Hence for any pure quasi-free state $\omega_{A}$  
the correlation matrix $A$ is of the form
$S^{T}S$ with $S \in Sp(E,\sigma)$ i.e.
$\omega_{A}=\omega_{I} \circ \alpha_{S}$ where $\omega_{I}$  
is called the Fock state. In other words any pure quasi-free state
can be obtained from the Fock state by a Bogoliubov transformation.
Because the Fock state is a state without correlations between
different modes it follows that in the state with the correlation
matrix $\omega_{A}$ the correlations are produced under the action
of the Bogoliubov automorphism  $\alpha_{S}$  ,  $S \in Sp(E,\sigma)$.
We can apply these considerations to the quasifree pure state 
$\widetilde \omega_{A}$ which arises by purification of the
quasifree state $\omega_{A}$. It follows that 
$\widetilde \omega_{A}$ can be obtained from the Fock state
by the Bogoliubov automorphism $\alpha_{S}$. 
As we have seen in the preeceding section the correlation
matrix of the state obtained by purification is given by:
\bigskip
\begin{equation}
\left(\matrix{S^T&0\cr 0&S^T\cr}\right)
\left(\matrix{\left(\matrix{D&0\cr0&D}\right)&
\left(\matrix{\sqrt{D^{2}-I}&0\cr0&\sqrt{D^{2}-I}\cr}\right)\cr
\left(\matrix{\sqrt{D^{2}-I}&0\cr0&\sqrt{D^{2}-I}\cr}\right)&
\left(\matrix{D&0\cr0&D\cr}\right)\cr}
\right)
\left(\matrix{S&0\cr 0&S\cr}\right)\\
\end{equation}
\bigskip
Hence the most general pure quasifree state is obtained
from the mixed quasifree state described by the
following correlation matrix :
\bigskip
\bea {\bf D}=
\left(\matrix{\left(\matrix{D&0\cr0&D}\right)&
\left(\matrix{\sqrt{D^{2}-I}&0\cr0&\sqrt{D^{2}-I}\cr}\right)\cr
\left(\matrix{\sqrt{D^{2}-I}&0\cr0&\sqrt{D^{2}-I}\cr}\right)&
\left(\matrix{D&0\cr0&D\cr}\right)\cr}
\right)
\eea
\bigskip
using a specific Bogoliubov transform which does not couple the
real and the fictious systems:

\bea \left(\matrix{S^T&0\cr 0&S^T\cr}\right)
\left(\matrix{J & 0 \cr 0 & -J \cr} \right)
\left(\matrix{S&0\cr 0&S\cr}\right)
= \left(\matrix{J & 0 \cr 0 & -J \cr} \right)
\eea
In the enlarged phase space there exists a Bogoliubov transform
${\bf S}$ such that ${\bf S}^T{\bf S}={\bf D}$.
If we denote by ${\bf J}= \left(\matrix{J&0\cr0&-J\cr}\right)$ 
then ${\bf S}^T{\bf J}{\bf S}={\bf J}$.
The solution is 
\bigskip
\bea {\bf S}=
\left(\matrix{\left(\matrix{\sqrt{{D+I \over \ 2}}&
0\cr0&\sqrt{{D+I \over 2}}\cr}\right)&
\left(\matrix{\sqrt{{D-I \over 2}}&
0\cr0&\sqrt{{D-I \over 2}}\cr}\right)\cr
\left(\matrix{\sqrt{{D-I \over 2}}&
0\cr0&\sqrt{{D-I \over 2}}\cr}\right)&
\left(\matrix{\sqrt{{D+I \over 2}}&0\cr0
&\sqrt{{D+I \over 2}}\cr}\right)\cr}
\right)
\eea
\bigskip
If we denote by ${\bf {\cal D}}$ the matrix
\bigskip
\bea
\left(\matrix{\left(\matrix{D&0\cr0&D}\right)&
\left(\matrix{0&0\cr0&0\cr}\right)\cr
\left(\matrix{0&0\cr0&0\cr}\right)&
\left(\matrix{D&0\cr0&D\cr}\right)\cr}
\right)
\eea
\bigskip
then the transition probability is given by

\bea  P(\omega_{A}, \omega_{B}) = (det({{\bf I}+ {\bf S}{\bf{\cal D}}
{\bf S}^T \over 2}))^
{-{1 \over 2}}  \eea  
and we reobtain the result (8.4).
\section{Conclusions}

The quasifree states are among the most simple and interesting,
from the physical point of view, because they are
states on which the quantum correlations
(the entanglement) between systems can be relatively easyly described.
We have applied to the finite dimensional systems the ideas and
the methods developed for the infinite dimensional systems 
by D. Kastler, J. Manuceau, A. Van Daele, A. Verbeure,
M. Fannes and A. S. Holevo to the case of finite dimensional
systems. For such systems, which are interesting from the
physical point of wiev, the theory was developed (with rare
exceptions) in an (apparently) independent way which is
based on noncommutative calculations.
In comparison, the methods developed for infinite dimensional
systems are more simple and more adequate in the finite
dimensional case. 

The fundamental object which contains
the entire information about a quasifree state is its
correlation matrix. We have shown in the present paper
how we can extract this information by elementary means.
For this purpose we have used results obtained many years
ago. Of great importance is a result of R. Balian, C. De Dominicis  
and C. Itzykson concerning the structure of covariance matrices.
According to this result all covariance matrices can be obtained
from the covariance matrix of a thermal state by orthogonal
symplectic transformations and squeezing symplectic transformations. 
This structure theorem is of fundamental importance and
explains all results on the structure of correlations
matrices obtained until now. All operations of restriction
to subsystems and of extension from a subsystem to the whole 
system can also be described in terms of correlation matrices.   
The transition probability from an arbitrary state to a squeezed
coherent state can be obtained as the value of the arbitrary state
on an element (a projector) of the CCR-algebra which is associated in a
natural way to any squeezed coherent state. In the present paper
we have shown only that the methods developed for squeezed states
can be applied also to the squeezed coherent states.
We have constructed explicitely the projector associated to any
squeezed coherent state. The transition probability is
an immediate result. 

The ultimate goal of these researches is
the desciption of the entanglement in the case of quasifree
states. We have shown that the most
entangled quasifree states are those obtained by the
purification of quasifree states. In the case of pure
states all corelations are obtained by Bogoliubov
transformations. For the states obtained by purifications
we have determined explicitely the Bogoliubov transformation
by which these states can be obtained from the Fock state.  
We have also determined explicitely the transition probability
from the state obatined by purification to the state of the
system obtained by direct product of the original system to
the twin fictitious one. This transition probability is a
decreasing function of the temperatures of the original
system. This means that a state obtained by the purification is
most correlated when the original state is most mixed. 
Hence we have obtained that the probability transition can
be considered as a measure of the entanglement.


\begin{references}
\bibitem{kas}
D. Kastler, Commun. Math. Phys {\bf 1}, 14 (1965).
\bibitem{man1}
J. Manuceau, Ann. Inst. H. Poincar\'e {\bf 8}, 139 (1968).
\bibitem{man2}
J. Manuceau and A. Verbeure, Commun. Math. Phys. {\bf 9}, 293 (1968).
\bibitem{van}
A. Van Daele  and A. Verbeure, Commun. Math. Phys. {\bf 20}, 268 (1971).
\bibitem{fan}
M. Fannes, Commun. Math. Phys. {\bf 51}, 55 (1976).
\bibitem{hol1}
A. S. Holevo, Problemy. Peredachi. Informacii {\bf 6}, 44 (1970).
\bibitem{hol2}
A. S. Holevo, Theoret. Mat. Fiz. {\bf 6}, 3 (1971).
\bibitem{hol3}
A. S. Holevo, Theoret. Mat. Fiz. {\bf 13}, 184 (1979).
\bibitem{hol4}
A. S. Holevo, IEEE Trans. Inform. Theory. {\bf IT-21}, 533 (1972).
\bibitem{scu1}
H. Scutaru, Phys. Lett. A  {\bf 141}, 223 (1989).
\bibitem{scu2}
H. Scutaru, Phys.. Lett. A {\bf 167}, 326 (1992).
\bibitem{alb}
P. M. Alberti and V. Heinemann, J. Math. Phys. {\bf 30}, 2083 (1989).
\bibitem{fol}
G. B. Folland,~{\it Harmonic analysis in phase space}~
( Princeton University Press, Princeton, 1989), page 177,
Proposition 4.22.
\bibitem{bal}
R. Balian, C. De Dominicis  
and C. Itzykson, Nuclear Physics {\bf 67}, 609 (1965).
\bibitem{bur}
D. J. C. Bures, Trans. Am. Math. Soc.  {\bf 135}, 199 (1969).
\bibitem{uhl}
A. Uhlmann, Rep. Math. Phys.  {\bf 9}, 273 (1976).
\bibitem{ara}
H. Araki and G. A. Raggio, Lett. Math. Phys. {\bf 6}, 237 (1982).
\bibitem{eza}
H. Ezawa, A. Mann, K. Nakamura
and M. Revzen, Ann. Phys. (N.Y.) 
{\bf 209}, 216 (1991).
\bibitem{hua}
H. Huang and G. S. Agarwal, Phys. Rev. A {\bf 49}, 52 (1994).
\bibitem{lang}
S. Lang, ${\it SL_{2} ({\bf R})}$~
(Addison-Wesley Publishing Company, Reading, 1975) Chapter XIV,
\S 1.
\end{references}
\end{document}